# A network analysis of Sibiu County, Romania

**Cristina-Nicol Grama**
Master in Tourism Management
University of Applied Sciences, Vienna, Austria
email: nicol.grama@gmail.com

**Rodolfo Baggio (\* corresponding author)**
Master in Economics and Tourism and Dondena Center for Research on Social Dynamics
Bocconi University
email: rodolfo.baggio@unibocconi.it

August, 2013

ABSTRACT

Network science methods have proved to be able to provide useful insights from both a theoretical and a practical point of view in that they can better inform governance policies in complex dynamic environments. The tourism research community has provided an increasing number of works that analyse destinations from a network science perspective. However, most of the studies refer to relatively small samples of actors and linkages. With this note we provide a full network study, although at a preliminary stage, that reports a complete analysis of a Romanian destination (Sibiu). Our intention is to increase the set of similar studies with the aim of supporting the investigations in structural and dynamical characteristics of tourism destinations.

INTRODUCTION

Modern network analysis methods are increasingly used in tourism studies and have shown to be able to provide scholars and practitioners with interesting outcomes. Nonetheless, the availability of investigations conducted at a broad scale on tourism destinations is still limited thus restraining our ability to understand the mechanisms that underlie the formation and the evolution of these complex adaptive systems. With this research note we aim at contributing to the field by augmenting the catalogue of tourism destination network studies and present the preliminary results of an investigation conducted in the county of Sibiu, a renowned Romanian destination.

THE DATA

Sibiu county lies in the heart of Romania (270 km from Bucharest) in the historical region of Transylvania. It is one of the most important tourism destinations of the country, a goal for almost every trip to Romania. In 2007, Sibiu has been the European Capital of Culture (together with Luxembourg). The destination accounts for roughly 250 000 arrivals and 460 000 overnight stays. Sibiu has a management organisation (AJTS) which is a public-private partnership in charge of promoting and marketing the county as a destination, and working in close collaboration with the local government. The tourism infrastructure is well developed and counts about 500 establishments providing more than 6000 rooms (a thorough description is in Richards & Rotariu, 2011).

The data for the network analysis were collected by using a number of publicly available documents (see Baggio et al., 2010 for details) complemented by a survey conducted on 551 operators (179 questionnaires were returned) aimed at validating the data collected and evaluating type and intensity of the relationships. The nodes of the network are the core tourism operators of the destination (i.e. accommodation, intermediaries, restaurants, travel agencies etc., as defined by UNWTO, 2000). Table 1 reports the distribution of the companies used for the study.

We estimate a 85%-90% completeness for the data collected, thus the metrics calculated can be reasonably considered as referring to the full network of Sibiu county. Here we present only a pure topological analysis, disregarding any attribute for the links and consider the network symmetric and unweighted. This, as known in the literature (Newman, 2010), is the basic and most important step for highlighting the structural characteristics of a complex network.

Table 1 Types of businesses included in the Sibiu destination network

| Type of business | % |
|---|---|
| Associations | 6.9% |
| Café-Bar-Pubs | 5.9% |
| Camping | 0.4% |
| Hotels | 7.9% |
| Motel/Hostel | 3.1% |
| Pensions | 52.5% |
| Private accommodation | 1.5% |
| Restaurants | 8.4% |
| Travel Agencies | 13.4% |

THE NETWORK ANALYSIS

*General results*

The network (Figure 1) comprises 551 nodes, with 14.5% disconnected elements. The isolated nodes were ignored for the quantitative analysis (i.e. the analysis was conducted on the strong connected component of the network). The main metrics calculated are shown in Table 2 (given the scope and the space restrictions for this note, we refer the reader to Baggio et al., 2010 or da Fontoura Costa et al., 2007 for a full description of the quantities reported).

The network is quite sparse (low density) but relatively compact (low average path length and diameter). It has also (proximity ratio) a marked small-world characteristic. All the local (nodal) measures look skewed in their distributions (see the difference between mean and median values). In the following paragraphs we discuss with more detail the most important results.

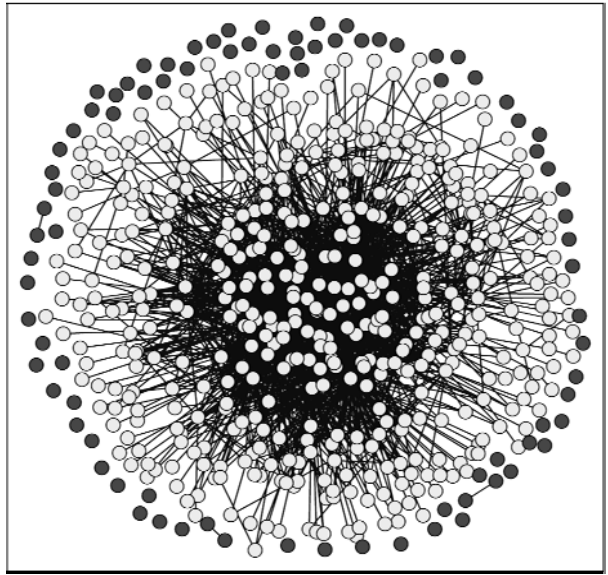

Figure 1 The Sibiu destination network (light coloured nodes are those belonging to the strong connected component used in the analysis)

Table 2 The main metrics calculated for the Sibiu destination network

| **Global metrics** | | **Local (nodal) metrics** | | |
|---|---|---|---|---|
| | | | *Mean* | *Median* |
| Nodes | 471 | | | |
| Edges | 2222 | Degree | 9.435 | 4 |
| Density | 0.020 | Clustering Coefficient | 0.325 | 0.269 |
| Diameter (max distance) | 7 | Path Length | 2.745 | 2.630 |
| Global efficiency | 0.386 | Betweenness | 0.004 | 4.57E-05 |
| Assortativity coefficient | -0.286 | Closeness | 0.014 | 0.001 |
| Proximity ratio | 18.585 | Eigenvector Centrality | 0.002 | 0.001 |
| Degree distribution exponent | 2.51±0.28 | Local efficiency | 0.426 | 0.482 |

*Degree distribution*

The statistical distribution of the degrees $k$ (number of links each node has) is an important parameter for a network and characterises its nature. In our case the degree distribution is consistent with a power-law functional form: $N(k) \sim k^{-\gamma}$. This, as known, is a typical signature of complexity for the system examined (Baggio, 2008; Newman, 2010). The exponent calculated is $\gamma = 2.51\pm0.28$ (calculations were performed following Clauset et al., 2009), Figure 2a shows the degree distribution and its cumulative version.

*Average neighbour connectivity and assortativity*

The degree distribution $N(k)$ accounts for the basic network topology but is not able to provide full information on its structural properties; networks with similar distributions can still exhibit different static or dynamic aspects. More information can be provided by looking for the existence of correlations between the degrees (Serrano et al., 2007). Two quantities can be used for this purpose: the distribution of the average degree of nearest neighbours $\langle K_{nn} \rangle$ and the Pearson correlation coefficient ($r$) between the nodal degrees (termed assortativity). Besides affecting dynamical processes such as the propagation of perturbations or the diffusion of information, these quantities are related to the capacity of the system to tolerate shocks without being disrupted but remaining able to recover in a reasonable period of time (resilience). A positive assortativity indicates a good resilience (the higher the better, see Serrano et al., 2007), while a negative one signals a possible high fragility of the system. Figure 2b shows a clear negative relationship between $\langle K_{nn} \rangle$ and $k$. This is further confirmed by the negativity of the coefficient: $r = -0.286$.

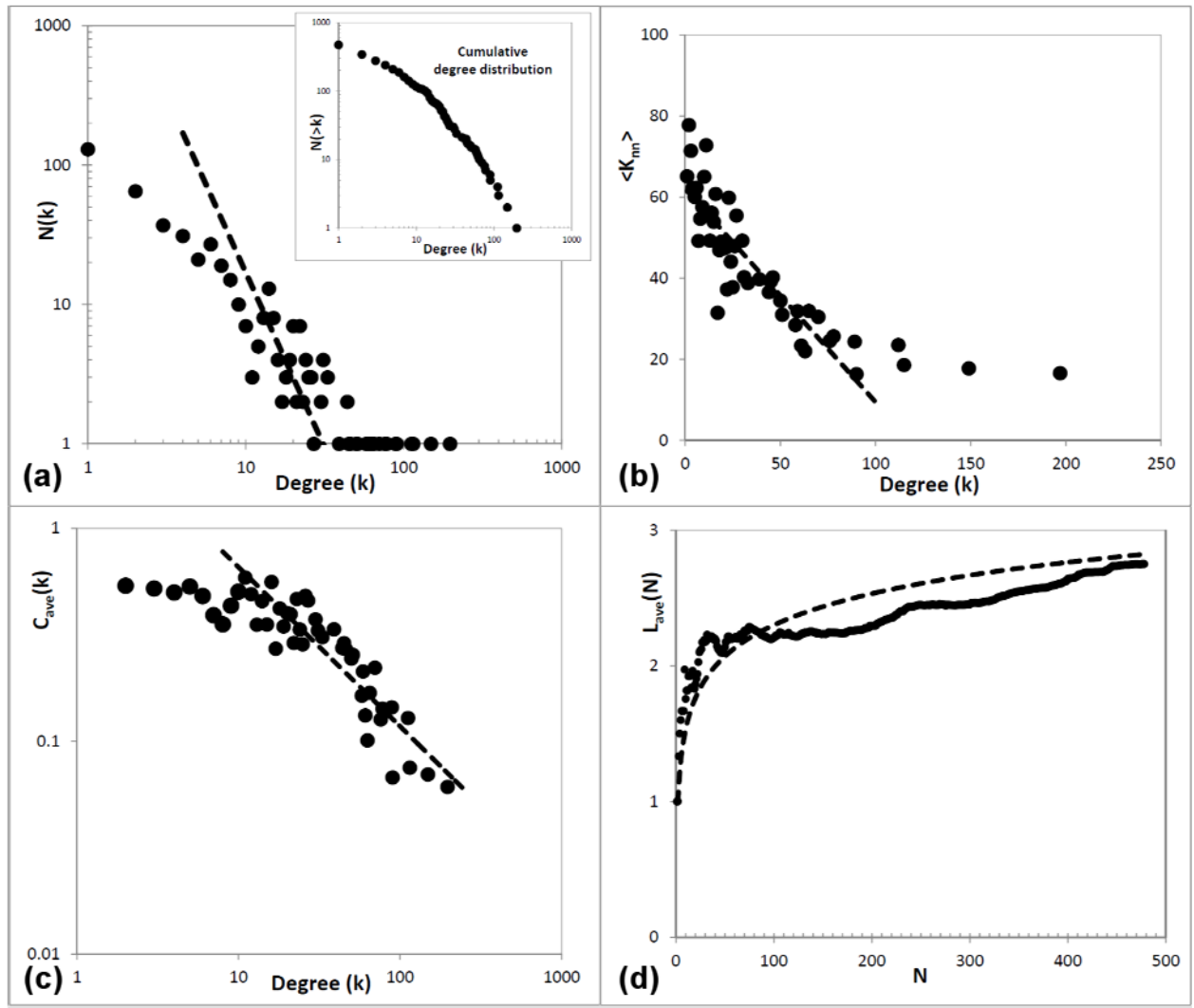

Figure 2 Results of the network analysis: a) degree distribution $N(k)$ with the power-law fit line, the inset shows the cumulative version $N(>k)$; b): the relationship between degree $k$ and average neighbour degree $\langle K_{nn} \rangle$ (dotted line is the OLS fit); c) distribution of average clustering coefficient as function of degree (dotted line is the power-law relationship); d): average path length as function of number of nodes (dotted line is a generic logarithmic function intended to guide the eye)

*Clustering coefficient*
The clustering coefficient $C$ measures the concentration of connections in the neighbourhood of a node and offers a measure of the heterogeneity of the local link density. The quantity can also be used as an indication of the extent to which the tourism stakeholders form cooperative communities inside the destination. Besides that, $C$, and its distribution as function of the nodal degrees, suggests a possible hierarchical organisation of the system (Ravasz & Barabási, 2003). This happens when the distribution of average clustering coefficients with respect to the degrees shows a power-law functional form: $C_{ave}(k) \sim k^{-\alpha}$. Figure 2c indicates that this holds for the main part of the distribution. The value for the exponent is: $\alpha = 0.75 \pm 0.24$.

*Average path length and small-world behaviour*
Watts and Strogatz (1998) have shown that a real network can exhibit a small-world behaviour, typified by a low average path length (distance between any two nodes) and a high clustering coefficient, while this does not happen in a network with links placed at random. In a small-world network the average path length increases logarithmically (or less) with the number of nodes N: $L_{ave}(N) \sim ln(N)$. Figure 2d shows clearly that this happens in our case.

*Modularity analysis*
Some systems see the presence of communities. Their elements are grouped based on some common characteristic such as type of business or geographical proximity. In a network, communities (modules) are groups of nodes that have denser connections between them than with nodes outside the group. They can be identified by using some stochastic algorithm that clusters the nodes according to their similarity connectivity (Fortunato, 2010). A quality index is usually defined (modularity index Q) that renders the goodness of the clustering. The index is always smaller than one, and can be negative when the network has no community structure. The analysis conducted on the Sibiu network (the algorithm of Clauset et al., 2004 was used) uncovered 13 modules with Q = 0.331, a relatively low value. The average size is of 37 nodes per module (although six of them have less than 10 nodes). Interestingly the most numerous communities include firms belonging (on the average) to six different types. This reconfirms the idea that the self-organising characteristics of the destination system produce cooperative groups that are quite different from those typically used, based on business type, when describing a tourism destination (see Baggio, 2011 for similar considerations).

CONCLUSIONS

The Sibiu destination looks to be a fragile complex system in which the stakeholders do not seem to be particularly inclined in forming cooperative groups, at least if of similar type. The limited modularity of the network, in fact, is due to communities made of heterogeneous actors. Even with these peculiarities, however, the network shows a good topological similarity with other destinations studied in the literature. The analysis presented here is a preliminary study. Future extensions will supply a deeper investigation by taking into account the type and intensity of the linkages uncovered in order to provide the destination management organisation with suggestions for a more effective governance of the system. For the time being this note contributes to the literature by augmenting the set of cases for which a full network analysis exists.